# A Photonic mm-Wave Local Oscillator

Robert Kimberk, Todd Hunter, C.-Y. Edward Tong, and Raymond Blundell

*Abstract*—A photonic millimeter wave local oscillator capable of producing two microwatts of radiated power at 224 GHz has been developed. The device was tested in one antenna of Smithsonian Institution's Submillimeter Array and was found to produce stable phase on multiple baselines. Graphical data is presented of correlator output phase and amplitude stability. A description of the system is given in both open and closed loop modes. A model is given which is used to predict the operational behavior. A novel method is presented to determine the safe operating point of the automated system.

## I. INTRODUCTION

A photonic local oscillator has been developed and used to pump a SIS mixer at 224 GHz. The local oscillator was installed in one antenna of the Submillimeter Array [1]. A single dish scan of Saturn was performed on February 24 2006. Observations of an ultra compact HII region and three quasars successfully produced stable fringes and amplitudes on a five element array. The test was performed on the evening of April 26 2006 UTC and the following day.

The configuration of the photonic lo not phase locked to an external reference signal (open loop) is shown in figure 1. The optical path consists of a single 1550 nm diode laser, a lithium niobate optical phase modulator, a Mach Zehnder interferometer and a photomixer whose output is connected

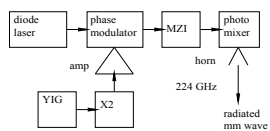

fig.1 open loop system

to a horn antenna. The photomixer was obtained from P.G. Huggard and B.N. Ellison of Rutherford Appleton Labs [2] in November of 2003. All optical devices and connections are polarization maintaining. The electrical path consists of a YIG synthesizer, a frequency doubler, and a power amplifier connected to the RF port of the phase modulator. The light from the laser is phase modulated at twice the YIG frequency, then converted to an amplitude modulated signal by the Mach Zehnder interferometer (MZI). The photomixer responds to the amplitude modulated signal and radiates that part of the am signal above the WR3 horn lower cutoff frequency. Figure 1 is a schematic of the open loop system.

## II. THE OUTPUT SPECTRUM

The voltage output of the photomixer, V, can be described by the following expression:

$$V = 1 + \Sigma\, J_n(2aK_p)\cos((\pi\omega_c / \omega_m) + n\omega_m t) \qquad (1)$$

where n ranges from plus to minus infinity, V is the amplitude of the output signal, $\omega_c$ is the optical angular frequency before modulation, and $\omega_m$ is the modulation angular frequency. $J_n(2aK_p)$ is the Bessel function of the first kind with $2aK_p$ as the argument. The quantity, $aK_p$, the peak phase deviation of $\omega_c$, is the product of the amplitude of the modulation signal, a, in volts and the optical phase modulator response, $K_p$, in radians / volt.

An examination of expression (1) shows that:

a) The photomixer output contains harmonics of the modulation frequency $\omega_m$.

b) Given $J_{-n} = -1^n J_n$ and $\cos(\theta) = \cos(-\theta)$, when $\pi\omega_c / \omega_m$ is an even multiple of $\pi/2$, odd harmonics cancel and only even harmonics appear at one output of the MZI. The other output of the MZI has the odd harmonics due to the extra $\pi$ phase shift at $\omega_c$.

c) Given $\cos(\theta + \pi/2) = -\cos(-\theta + \pi/2)$, when $\pi\omega_c / \omega_m$ is an odd multiple of $\pi/2$ then the even harmonics cancel and only odd harmonics appear at the output.

d) When $\pi\omega_c / \omega_m$ meet the conditions of b) or c) the amplitudes of the harmonics are determined by $2aK_p$. Increasing the power into the optical phase modulator RF port will increase the number of harmonics.

e) For values of $\pi\omega_c / \omega_m$ that do not meet the conditions of b) or c) both odd and even harmonics exist at both MZI outputs.

The frequency of the laser, $\omega_c$, is controlled by the laser temperature. The laser used, JDS CQF935/808, has a wavelength temperature tunability of about 0.1 nm / °C, equivalent to about $7.85 * 10^{10}$ rad /sec / °C at 1550 nm. Figure 4 shows the measured power of a selected harmonic as the laser temperature is varied. Note that the power output completes a cycle in about 3° C. This corresponds to about $23.55 * 10^{11}$ rad / sec, about $\omega_m$. Note also that the maximum output power of the harmonic occurs at the middle value of photomixer bias current.

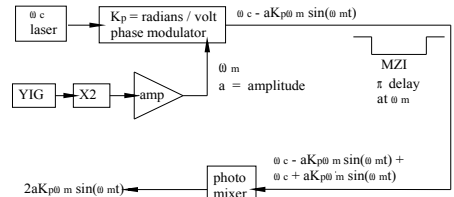

Fig. 2. output as difference of phase modulated laser and delayed phase modulated laser

Expression 1 assumes a photomixer with a flat response over the output frequency. To estimate the true output power the results of expression 1 need to be multiplied by the photomixer's frequency response.

We normally tune the laser temperature so that the two outputs of the MZI are purely even and odd harmonics of the modulation frequency. The two advantages of this mode are that the frequency separation of harmonics is greatest (two times the modulation frequency), and the power is maximized for any harmonic.

Figure 2 shows the instantaneous angular frequency [3] (d θ/dt , where θ = phase) at key points along the signal path. Instantaneous frequency is a useful concept when dealing with frequency or phase modulation. The MZI splits the incoming power, phase shifts one component, then sums the two components. The greatest frequency multiplication occurs when the modulation frequency is equal to one half of the free spectral range (FSR) of the MZI. This condition adds a π radian phase delay at the modulation frequency. The instantaneous difference frequency then becomes:

$2aK_p \omega_m \sin(\omega_m t)$
$= \omega_c + aK_p \omega_m \sin(\omega_m t) - (\omega_c + aK_p \omega_m \sin(\omega_m t + \pi))$
$= \omega_c + aK_p \omega_m \sin(\omega_m t) - (\omega_c - aK_p \omega_m \sin(\omega_m t))$.

The resulting instantaneous frequency, $2aK_p \omega_m \sin(\omega_m t)$, describes a sinusoidal chirp. The maximum frequency of the

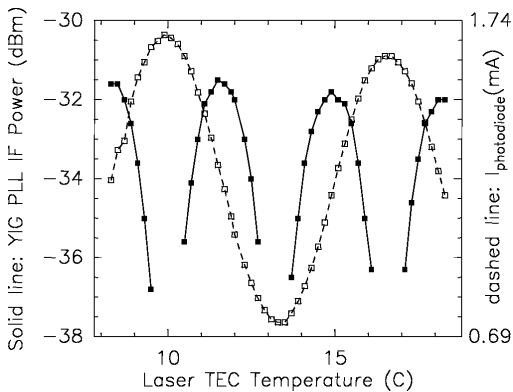

Fig.4 laser temperature vs single tone power and photomixer bias current chirp is equal to twice the product of the peak phase deviation, $2aK_p$, and the modulation frequency, $\omega_m$. As the modulation frequency is moved away from FSR/2 the maximum instantaneous frequency of the output is reduced.

### III. CLOSED LOOP SYSTEM

In order to operate a radio interferometer, the local oscillators of all the receivers must be phase locked to a common reference tone. The closed loop configuration of the photonic local oscillator, where the phase of the mm wave output is locked to a reference frequency, is shown in figure 3. A second photomixer is connected to the unused MZI output. The output of the added photomixer is connected to a harmonic mixer, which in turn is connected to the IF port of the phase locked loop (PLL) through a frequency diplexer and IF amplifier. The PLL control output is connected to the YIG fine tune input, and controls the YIG output phase. In this way the YIG is locked to a down converted 112GHz, the third multiple of $\omega_m$. The multiples of the modulation frequency, $\omega_m$, are all related with respect to phase. Locking to any multiple stabilizes the phase of all the other multiples. Not shown in figure 3, is the 6 to 8 GHz lo pumping the harmonic mixer, the 109 MHz reference connected to the PLL, and the coarse tuning signal to the YIG.

### IV. OPEN LOOP PHASE NOISE AND STABILITY

The open loop phase drift of the system was tested by

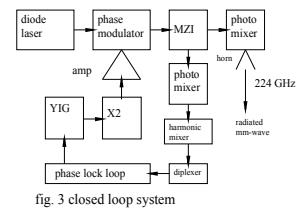

fig. 3 closed loop system

power splitting the output of a microwave synthesizer, modulating our system with one of the two identical outputs and using the other output to pump a conventional multiplier chain. The synthesizer's output frequency was 12.5 GHz and the final frequency in both legs was 75 GHz. The outputs of both the photonic and conventional multipliers were down converted using mixers with a common local oscillator and the phase was compared with a vector voltmeter. Differential phase drifts of 5 degrees / hr were observed. The phase drift tracked the room temperature change. The differing lengths and temperature coefficients of coaxial cable and optical fiber involved in this test might explain the observed phase drifts. One degree at 75GHz is equal to about 7.4μm in quartz fiber.

The phase noise of the output of the photonic local oscillator is dominated by the multiplied phase noise of the modulation source. We expect the system to act like a classical multiplier, the output phase noise will be the phase noise of modulation source plus 20 log M. M is equal to the ratio output frequency / input frequency.

The open loop phase noise of a heterodyne two laser local oscillator is similar to the phase noise of two lasers. In contrast the contribution of the single laser phase noise to the open loop phase noise of our system is determined by phase noise of the laser times $(\tau/\tau_c)^2$, where $\tau_c$ is the coherence time of the laser and $\tau$ is the path delay of the MZI. The value of $(\tau/\tau_c)^2$ is very small for a MZI with a free spectral range of 75 GHz (a delay of $1.3*10^{-11}$s) and a laser coherence time of $10^{-6}$ s.

## V. ASTRONOMICAL OBSERVATIONS

The photonic local oscillator was installed in antenna 6 of the Submillimeter Array on Mauna Kea, Hawaii, and tuned to 224 GHz. An observation of the quasar 3C273 was obtained on the evening of April 26 2006 UTC. The output of the correlator yielded stable phase and amplitudes for each of the baselines
with the seven operating antennas. The next day observations of the ultra compact HII region G138.295+1.555 [4], the quasar 3C84, and the quasar 3C454.3 were obtained with a five element array. Figure 5 shows the phase (dots) and amplitude (gray trace) of each

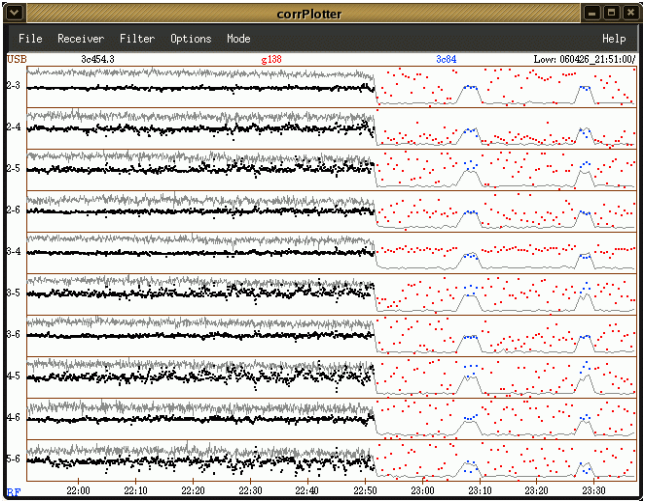

Fig.5 phase and amplitude at correlator output. The sources are 3C454.3, G138.295+1.555, and 3C84

baseline at the correlator output. Baselines that include antenna 6 do not show a greater scatter of values than baselines that do not include antenna 6.
Baseline lengths range from 16 m at the top to 69 m at the bottom of figure 5. Line spectra of 13CO2-1 and 12CO2-1 were observed in the source G138.295+1.555. The 232.4 GHz beacon attached to the Subaru building was observed as shown in figure 6. Figure 7 is a graph of the 12CO2-1 line from G138.295+1.555 at the correlator ouput.

## VI. AUTOMATION

The photomixer bias current limit is 5mA, above which the photomixer will be damaged. The bias current is a function of laser power and temperature as well as modulation frequency. This fact makes it necessary to find a path through the various combinations of the variables that does not destroy the photomixer. To explore the effect of temperature on the modulated output power, we biased the laser with 60 milliamps of current (about 1/8 the normal operational current) and measured the photodiode current as a function of laser temperature at several modulation frequencies. As seen in Figure 4, we find a sinusoidal behavior of photodiode current with respect to laser temperature. Although the amplitude of the sinusoid depends on modulation frequency, we discovered that there exist "fixed point" temperatures where the photodiode current is independent of modulation frequency.
We next locked the photonic LO using a W-band harmonic mixer and a digital PLL. We measured the power of the locked signal at the PLL IF monitor port as a function of laser temperature. Figure 4 shows that the IF power is at maximum when the laser is operated at the fixed point temperatures.
To optimize the IF power, we can use the following algorithm. Change the modulation frequency by some significant amount (1 GHz) and observe the change in the photomixer current. Adjust the laser temperature slightly upward and change the modulation frequency back to the initial value. If the photomixer current decreases, then continue adjusting the laser temperature upward until the photomixer remains constant with respect to the modulation frequency. Conversely, if the photomixer current increases, then adjust the laser temperature downward instead.

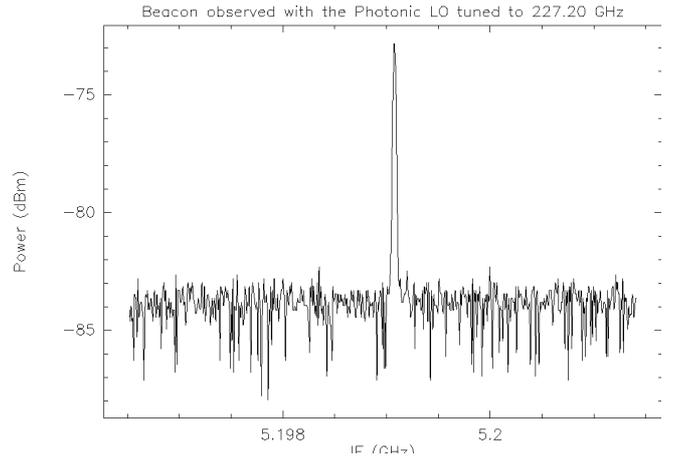

Fig.6 Beacon on Subaru Building observed from SMA antenna 6

## VII. CONCLUSION

We have demonstrated that a photonic local oscillator can be sufficiently phase and amplitude stable to be used as part of a radio interferometer at mm wave lengths. We believe that an increase in the frequency of modulation will allow for greater output frequencies as well as greater separation between the harmonics of the modulation frequency.

## VIII. Acknowledgment

The authors want to thank Smithsonian Institution for its support. We want to thank Bob Wilson for early insight and for his belief that the system would work when most were skeptical. We want to thank Hugh Gibson for a numerical analysis of the system with both the Mach Zehnder and Fabry Perot interferometers, and Robert Christensen for instrumental instruction and observational guidance at Mauna Kea. As well we want to acknowledge those people that developed the concept before us, O.I. Kotov et al [5], E.J. Bochove et al [6] to name a few. A development of the ideas in this paper in greater detail was published in an earlier ISST proceedings [7] and may be a useful background.

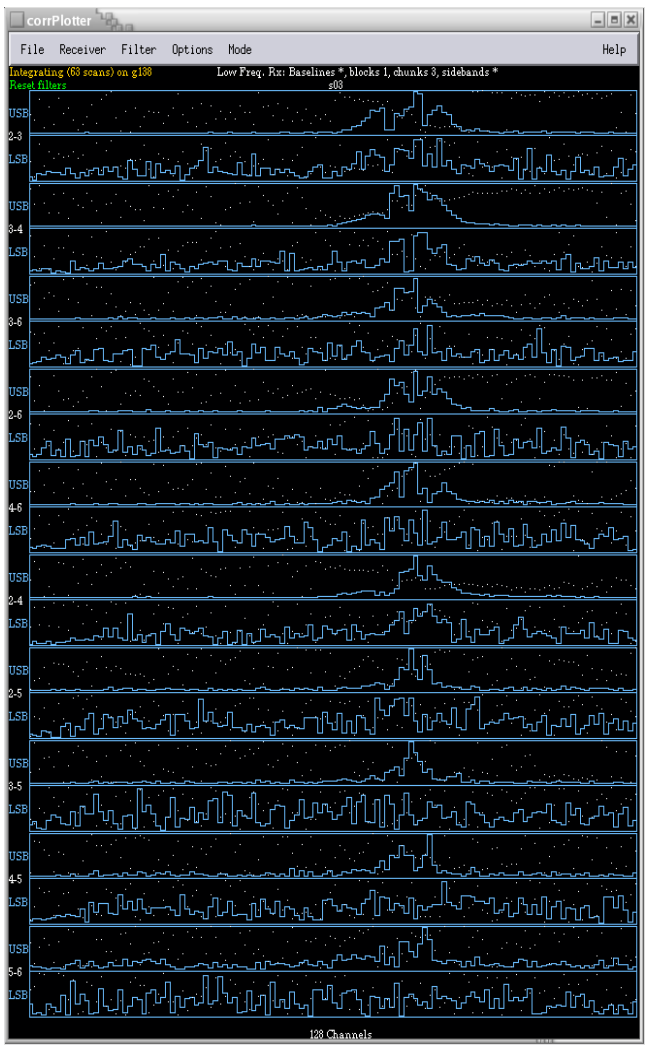

Fig.7 12CO2-1 spectral line observed in source G138.295+1.555